\documentclass{emulateapj}
\shorttitle{Formation of Saturn in 3.4 Myr}
\shortauthors{Dodson-Robinson et al.}
\begin{document}

\title{Saturn Forms by Core Accretion in 3.4 Myr}

\author{Sarah E. Dodson-Robinson\altaffilmark{1,2},
Peter Bodenheimer\altaffilmark{3},
Gregory Laughlin\altaffilmark{3},
Karen Willacy\altaffilmark{4},
Neal J. Turner\altaffilmark{4},
C. A. Beichman\altaffilmark{2}}

\altaffiltext{1}{Formerly Sarah E. Robinson}

\altaffiltext{2}{NASA Exoplanet Science Institute, California Institute
of Technology, 770 South Wilson Avenue, Pasadena, CA 91125;
sdr@ipac.caltech.edu, chas@ipac.caltech.edu}

\altaffiltext{3}{University of California Observatories/Lick
Observatory, Department of Astronomy and Astrophysics, University of
California at Santa Cruz, Santa Cruz, CA 95064; peter@ucolick.org,
laughlin@ucolick.org}

\altaffiltext{4}{Jet Propulsion Laboratory, California Institute of
Technology, Pasadena, CA 91109; karen.willacy@jpl.nasa.gov,
neal.turner@jpl.nasa.gov}

\begin{abstract}

We present two new {\it in situ} core accretion simulations of Saturn
with planet formation timescales of 3.37~Myr (model S0) and 3.48~Myr
(model S1), consistent with observed protostellar disk lifetimes.  In
model S0, we assume rapid grain settling reduces opacity due to grains
from full interstellar values (Podolak 2003).  In model S1, we do not
invoke grain settling, instead assigning full interstellar opacities to
grains in the envelope.  Surprisingly, the two models produce nearly
identical formation timescales and core/atmosphere mass ratios.  We
therefore observe a new manifestation of core accretion theory: at large
heliocentric distances, the solid core growth rate (limited by Keplerian
orbital velocity) controls the planet formation timescale.  We argue
that this paradigm should apply to Uranus and Neptune as well.

\end{abstract}




\section{Introduction}
\label{introduction}

The discrepancy between the observed lifetimes of protostellar disks
(2--3~Myr; Haisch et al. 2001) and the length of time required for
planet formation by core accretion ($> 8$~Myr; Pollack et al. 1996) has
long presented a problem for planet formation theory.  However, in 2005,
two new models of planet formation showed that the core accretion-gas
capture process could form gas giants within 2.5~Myr.  Hubickyj et al.
(2005) modeled Jupiter's formation {\it in situ} at 5.2~AU and found
that Jupiter could grow from a $0.1 M_{\oplus}$ core to its present mass
in 2.2~Myr.  The core accretion models of Alibert et al. (2005) allowed
both Jupiter and Saturn to form concurrently within 2.5~Myr in disk of
mass $0.035 \: M_{\odot} < M_{\rm disk} < 0.05 \: M_{\odot}$.

Each model employed a different approach in order to speed up giant
planet formation.  Alibert et al. added Type I migration to the core
accretion model: the inward motion of the protoplanets allows them to
receive a fresh supply of planetesimals and gas as they move into
undepleted regions of the solar nebula.  The accretion rate is then no
longer limited by the rate at which the protoplanet's Hill sphere
expands, so the giant planet formation timescale decreases by up to an
order of magnitude.  These models require proto-Jupiter to have an
initial semimajor axis of $a \geq 9.2$~AU and proto-Saturn to begin
forming at 11.9~AU, and migrate to their current positions on a $\sim
2$~Myr timescale.


One notable feature of the \cite{alibert05} core accretion model is that
it successfully predicts the heavy metal content of Jupiter and Saturn's
atmospheres according the clathrate hydrate trapping theory of Lunine \&
Stevenson (1985).  However, the Type I migration rate is a free
parameter: Alibert et al. note that the analytical work of
\cite{tanaka02} predicts migration rates far too large to be consistent
with the observed frequency of extrasolar planets.  The authors get
around this problem by scaling the \cite{tanaka02} by an arbitrary
factor $f_1$, where $0 \le f_1 \le 0.03$.  Finally, there is one more
basic assumption underlying the model, which is that the gas/solid ratio
beyond the ice line is constant at $G/S = 70$.


The {\it in situ} planet formation models of Hubickyj et al. (2005)
decrease Jupiter's formation time by requiring that grains quickly
settle to the bottom of the protoplanetary envelope, where they are
destroyed by sublimation (Podolak 2003).  Assuming grain settling lowers
envelope opacity due to grains to $\sim 2$\% of the interstellar value,
the gas can contract efficiently and make way for new material entering
the protoplanet's Hill sphere.  The most important free parameter in the
Hubickyj et al. models is the solid surface density of planetesimals in
the planet's feeding zone: since the protoplanet doesn't move through
the disk, it requires a feeding zone with $\Sigma_{\rm solid} \ga
10$~g~cm$^{-2}$ in order to attain the $\sim 15 M_{\oplus}$ core
necessary for accreting a massive gaseous envelope (Papaloizou \& Nelson
2005).

The availability of new calculations of solid surface density as a
function of heliocentric distance in the solar nebula (Robinson et al.
2008; hereafter Paper 1) raises the possibility of extending {\it in
situ} core accretion simulations to include Saturn.  Since the Nice
model of Tsiganis et al. (2005) predicts that proto-Saturn migrated 1~AU
outward at most, we consider the {\it in situ} approximation reasonable
when applied to Saturn.  By providing theoretically and observationally
motivated values for solid surface density $\Sigma_{\rm solid}$, the
Paper 1 results move core accretion simulations away from parameter
studies and toward determinism.

Since the solar nebula had to be capable of forming both Jupiter and
Saturn concurrently (and of course Uranus and Neptune, the formation of
which we will examine in future work), we first assess the ability of
the Paper 1 solar nebula model to produce Jupiter.  In Paper 1, the
value for $\Sigma_{\rm solid}$ after $10^5$~yr of solar nebula evolution
is 13.2~g~cm$^{-2}$.  Adopting the relationship between solid surface
density and Jupiter formation time at 5~AU calculated by Robinson et al.
(2006), 
\begin{equation}
{t_{\rm form} \over 1 {\rm Myr}} = \left ({\Sigma_{\rm solid} \over 25.0
\: {\rm g \: cm}^{-2} } \right )^{-1.44} ,
\label{tjup}
\end{equation}
we find that the Paper 1 results allow Jupiter to form in 2.5~Myr.
Built in to the \cite{robinson06} scaling relation is the assumption
that efficient grain settling leads to protoplanetary envelope opacities
of $\sim 2\%$ those of interstellar grains.  In this Letter, we will
relax this assumption and also investigate the limiting case of 100\%
interstellar grain opacity with respect to Saturn.

If the fiducial disk from Paper 1 can form Saturn within 2--3~Myr, we
will have successful core accretion models of the two gas giants forming
near their present positions in a gravitationally stable disk (see Paper
1 for a discussion of the solar nebula dynamics).


In \S \ref{methods}, we describe our theoretical treatment of the core
accretion process.  In \S \ref{results}, we discuss the results of our
simulations, with special emphasis on formation timescale, effect of
atmospheric opacity, and core/atmosphere mass ratio.  We present our
conclusions in \S \ref{caconclude}.

\section{Theoretical Treatment of Core Accretion}
\label{methods}

We use the theoretical model of planet formation described by Laughlin
et al. (2004) to model the core accretion and gas capture of
proto-Saturn.  Initially, a protoplanetary core of mass $M_{\oplus}$ is
embedded at Saturn's heliocentric distance, 9.5~AU, in a viscously
evolving disk of age $1.5 \times 10^5$~yr, surrounding a T-Tauri star of
mass $1 M_{\odot}$.  We assume that by $1.5 \times 10^5$~yr, the
available dust has formed 100~km planetesimals that are invulnerable to
gas drag (Weidenschilling 1977): planetesimal orbits are modified only
by interactions with proto-Saturn.  Gas temperature and density
are regulated by viscous evolution of the solar nebula.  We use the
time-evolving temperature and density at 9.5~AU, beginning at $t = 1.5
\times 10^5$~yr, as calculated in Paper 1.


The contraction and buildup of protoplanetary cores and their gaseous
envelopes embedded in our model evolving disk are computed with a
Henyey-type code (Henyey et al. 1964).  Following the argument of
\cite{podolak03} that grain settling in the protoplanetary envelope
would reduce envelope opacity where grains exist, we adopt grain
opacities of 2\% of the interstellar values used in \cite{pollack96} in
our fiducial model, which we will call S0.  However, in order to assess
the effect of envelope opacity on Saturn's formation timescale, we
present a second core-accretion simulation, S1, using full interstellar
grain opacity.

We use a core accretion rate of the form
\begin{equation}
{dM_{\rm core} \over dt} = C_1 \pi \Sigma_{\rm solid} R_c R_h \Omega
\label{coregrowth}
\end{equation}
(Papaloizou \& Terquem 1999), where $\Sigma_{\rm solid}$ is the surface
density of solid material in the disk, $\Omega$ is the orbital frequency
at 9.5 AU, $R_c$ is the effective capture radius of the protoplanet for
solid particles, $R_h = a[M_{\rm planet} / (3 M_*)]^{1/3}$ is the tidal
radius of the protoplanet (where $a$ is the semimajor axis of the
protoplanet's orbit), and $C_1$ is a constant near unity.

The outer boundary conditions for the protoplanet include the decrease
with time in the background nebular density and temperature.  During the
late phase of planet growth, when planetesimals may be ablated by the
massive envelope before reaching the core, we consider a planetesimal
captured if it deposits 50\% or more of its mass in the envelope.  At
this stage, we invoke the sinking approximation and assume the ablated
planetesimal debris sinks rapidly to the planet core without leaving
remnants in the envelope.

\section{Results}
\label{results}

In both simulations (S0 and S1), we start the core accretion model with
midplane temperature, gas density and solid surface density from the
solar nebula model of Paper 1.  This model has two key features favoring
planet formation that are missing from passive disk models: (1) viscous
stresses drive the initial $\Sigma \propto R^{-3/2}$ surface density
profile toward uniformity, so that Saturn's feeding zone gains mass
during the first $5 \times 10^4$~yr of disk evolution, and (2) the
presence of hydrated ammonia ice at the snow line increases the solid
surface density by 7\% over the standard water ice--rock--refractory
CHON mixture.  We use a starting solid surface density of
8.6~g~cm$^{-2}$ (Paper 1) which decreases with time as proto-Saturn
captures planetesimals.


Figure 1a shows the growth of Saturn from a core of $1 M_{\oplus}$ to
its present-day mass of $95 M_{\oplus}$.  Solid lines correspond to the
S0 model (in which grains quickly settle and sublimate, reducing their
contribution to envelope opacity by an assumed factor of 50), and dashed
lines show the S1 model (in which grains stay in the envelope and
opacity due to grains takes on the full interstellar value).  An
important property of our model, and one of two key results of this
Letter, is Saturn's formation time: the planet attains its current mass
in only 3.37~Myr for model S0 and 3.47~Myr for model S1.  These are the
first {\it in situ} core accretion models of Saturn with formation times
within $1 \, \sigma$ of observed protostellar disk lifetimes.

The second key result is that Saturn's formation time is nearly
independent of the assumed grain opacity in the envelope.  This
surprising result occurs because Saturn's core growth rate is limited by
the Keplerian speed in the feeding zone: a slow-moving core takes a long
time to encounter and capture planetesimals (see Equation
\ref{coregrowth}), and may never reach isolation mass.  Whereas during
Jupiter's formation early core buildup takes only 0.5~Myr and the gas
contraction phase dominates planet growth (Hubickyj et al. 2005), in
Saturn's case, core growth lasts until the planet reaches its final
mass.  Only in the last $6 \times 10^5$~yr of growth does Saturn possess
a gas mass of $> 1 M_{\oplus}$.

Given that the planet has a negligible envelope mass throughout most of
its formation, it stands to reason that envelope opacity would not exert
much influence on Saturn's formation timescale.  The idea that solid
growth rate controls Saturn's formation is consistent with the planet's
high core/atmosphere mass ratio: $9-23\%$, as opposed to $< 3\%$ for
Jupiter (Saumon \& Guillot 2004).


\subsection{Core/Atmosphere Mass Ratio}

If all accreted planetesimals and their ablated debris end up in the
solid core, as is assumed in our model, Saturn's total core mass reaches
$44 M_{\oplus}$ in model S0 and $54 M_{\oplus}$ in model S1.  Based on
gravitational moment measurements and internal structure modeling,
Saumon \& Guillot (2004) place Saturn's core at 9--22$M_{\oplus}$ and
total heavy element content at 13--28$M_{\oplus}$.  Our model Saturn has
a heavy element mass that is too high by at least a factor of 1.6 (model
S0).  We note, however, that the carbon enrichment in Saturn's
atmosphere has recently been revised upward to $(C/H) / (C/H)_{\odot} =
7$ (Flasar et al. 2005), which may allow for a larger heavy element
inventory than previously thought.

One method of reducing the core mass, though not the total heavy element
mass, is to account for planetesimal disruption not only by ablation of
debris (\S \ref{methods}), but by sublimation of volatiles.
Our planetesimals are 50\% H$_2$O by mass, 6\% NH$_3$, and
1\% other ices, such as HCN and H$_2$S, for a total of 57\% ice.  These
ices could sublimate either during infall through the envelope,
analogously to meteorites in Earth's atmosphere, or upon impact with
the solid core.  Helled et al. (2008) found that sublimation of
volatiles during the collapse of a giant planet formed by disk
instability (Boss 2005) is efficient: refractory silicate grains
sediment to form a core, while ices remain in the planet atmosphere.


Tingle et al. (1991) tested the survival of volatiles experiencing
hypervelocity impacts ($v \sim 1 \: {\rm km \: s}^{-1}$) by shocking
samples of the Murchison meteorite with pressures up to 36~GPa.  They
found that 70\% of organic and organosulfuric material, including
H$_2$S, sublimates upon experiencing an impact with $v > 1.5 \: {\rm km
\: s}^{-1}$.  The volatiles in Saturn-building planetesimals are not
likely to survive a high-velocity impact and subsequent contact with the
hot protoplanetary core, $T \geq 3000$~K, in solid form.  If we assume
all accreted ices undergo a phase transition from solid to gas, Saturn's
core mass drops to $19 M_{\oplus}$ (S0) and $23 M_{\oplus}$ (S1), which
are near the range determined by Saumon \& Guillot (2004).  However, the
total heavy element/hydrogen mass ratio in the planet is still higher
than observed: Saturn is 50--70\% hydrogen by mass, whereas our model
predicts a hydrogen mass fraction of only 42\% (S0) and 31\% (S1).

Another possible way to reduce Saturn's heavy element mass without
slowing the planet's growth is to cut off solid accretion midway
through planet formation.  This approach simulates the effect of another
embryo competing for planetesimals.  Although our solid accretion rates
are calculated for the monarchic growth paradigm, in which planetesimal
dynamics are determined by proto-Saturn only, this scenario is an
approximation: Kokubo \& Ida (1998) predict an oligarchic planet
formation epoch with competing embryos spaced $\sim 10$ Hill radii apart
that lasts for $\sim 1$~Myr.

Hubickyj et al. (2005) tested the effect of a core accretion cutoff on
Jupiter's formation and found that as long as the core has mass $\geq 10
M_{\oplus}$ before solid accretion ceases, the planet reaches
hydrodynamic gas accretion even more quickly than when solid accretion
continues unchecked: 0.78~Myr vs. 2.22~Myr.  Continuous, late-stage
planetesimal accretion slows planet formation by depositing kinetic
energy in the protoplanetary envelope and inhibiting gas contraction.

The competing embryo scenario holds promise for bringing Saturn's
core/atmosphere mass ratio into agreement with observations while still
retaining the quick formation time.  However, it is a double-edged
sword: oligarchs with overlapping zones of gravitational influence can
increase the RMS planetesimal eccentricity, $\langle e^2 \rangle^{1/2}$,
far more efficiently than a single monarch core, decreasing the
gravitational focusing ability of all embryos.  \cite{thommes03} and
\cite{ida93} find that significant planetesimal stirring can occur when
embryos are between $10^{-5}$ and $10^{-2} \: M_{\oplus}$.  Indeed,
\cite{fortier07} replaced Equation \ref{coregrowth} with the analytical
oligarchic growth rate of \cite{ida93} in {\it in situ} core
accretion simulations of Jupiter and found formation times of 10-20~Myr.

Thommes et al. (2008) created a self-consistent planet formation model
that included gas disk evolution, planet-disk interactions (including
gap opening and gas accretion onto solid cores), and planet-planet
interactions.  Gas giants with Jupiter-like core/atmosphere mass ratios
consistently emerged in disks with mass $M_{\rm disk} \ga 0.06
M_{\odot}$.  Even a modest amount of migration, as occurred in the Solar
System, appears to compensate for the inefficiency of {\it in situ}
oligarchic growth.  The true Saturn formation scenario probably involved
several $\sim 1 M_{\oplus}$ embryos spanning 9-12~AU, some outward
motion of proto-Saturn, and possible competition with Neptune, forming
near 12~AU (Tsiganis et al. 2005).


\subsection{Isolation Mass}

One new feature of both models, S0 and S1, is that Saturn does not need
to accrete all the planetesimals in its feeding zone to reach its
current mass.  Figure 1b shows the remaining solid surface density, not
incorporated in Saturn's core, as a function of time.  (The upturn near
the end of the simulation results from the rapid expansion of the
planet's feeding zone during hydrodynamic gas accretion and the
assumption that planetesimals are always uniformly distributed, which is
unphysical if the planet grows more quickly than planetesimal
redistribution can occur.)  In both models, there are still
2~g~cm$^{-2}$ of solids left in Saturn's feeding zone at $t = 3.4$~Myr.
This means the core never reaches isolation mass.  Following Lissauer
(1993), we calculate
\begin{equation}
M_{\rm iso} = 0.0021 \, \Sigma_{\rm solid}^{3/2} \, a^3 = 58 M_{\oplus}
,
\label{miso}
\end{equation}
whereas Saturn's core attains a mass of $44 M_{\oplus}$ in model S0 and
$54 M_{\oplus}$ in model S1.

Both Hubickyj et al. (2005) and Pollack et al. (1996) found that giant
planets pass through a lengthy plateau phase in which both solid and gas
accretion rates are low and planet mass changes very little with time.
This phase begins when the planet core nears isolation mass, and its
duration is regulated by gas contraction efficiency.  Examining Figure
1a, we see that Saturn never experiences this mass plateau: both solid
and gas accretion rates increase with time.

With extra solid and gas mass remaining in the disk after 3.4~Myr, why
would Saturn not continue to grow?  We arbitrarily stop the simulation
once Saturn reaches $95 M_{\oplus}$, but our disk could quickly form a
more massive planet.  Explaining Saturn's low mass in the context of the
disk models in Paper 1 may require either another nearby embryo
competing for both solids and gas--perhaps Neptune, forming near 12.5 AU
according to the Nice model (Tsiganis et al. 2005)---or a mechanism for
disk dissipation that begins after a few Myr.  Photoevaporation
(Alexander et al. 2006) is one possibility: the disk near 10~AU begins
to be disrupted by ionizing radiation at 2--3~Myr.  We also require a
mechanism to populate the Oort cloud with $40 M_{\oplus}$ of comets
(Weissman 1996).  Saturn scattering the $4-14 M_{\oplus}$ it does not
accrete into the Oort cloud would be a good start, and Jupiter could
contribute more material during the late stages of its formation and/or
migration.

Is the concept of isolation mass relevant for any giant planet except
Jupiter?  From Equation \ref{miso}, we see that $M_{\rm iso}$ increases
with radius if $\alpha \leq 2$ in the power-law surface density profile
$\Sigma \propto R^{- \alpha}$.  Extrapolating from the simulation
presented here and the near-flat solid surface density profile seen in
Paper 1, we propose that Saturn, Uranus and Neptune do not reach
isolation mass.  Due to their low Keplerian orbital speed, planet cores
in the outer solar nebula experience a lower planetesimal collision rate
than proto-Jupiter.  Instead of their formation timescale being governed
by gas contraction efficiency, the solid accretion rate, mediated by
$\Omega$ and $\Sigma$, is the critical factor.  The hypothesis that the
three outer planets never reach isolation mass is consistent with the
planets' bulk composition: all three have solid/gas ratios far higher
than Jupiter.


\section{Conclusions}
\label{caconclude}

Based on the solid surface densities calculated in Paper 1, we have
created core accretion models of Saturn with a formation timescale of
3.37-3.48~Myr, which is compatible with observed protostellar disk
lifetimes.  Unlike previous studies of Jupiter (e.g. Hubickyj et al.
[2005]), we find that grain opacity in the protoplanetary envelope has
virtually no effect on Saturn's formation timescale.  The same solar
nebula model that was the basis for simulations S0 and S1 is capable of
forming Jupiter in 2.5~Myr, assuming reduced grain opacities.



Finally, this is the first {\it in situ} core accretion model of a gas
giant that lacks a plateau phase, in which the planet's mass remains
nearly constant at approximately the isolation mass.  We postulate that
the low orbital speeds far from the sun prevented Saturn, Uranus and
Neptune from accreting solids efficiently enough to reach isolation
mass.  These planets never experienced a gas-only accretion phase,
as did Jupiter, and consequently have much higher core/atmosphere mass
ratios.


We thank Geoff Bryden and Jonathan Fortney for helpful conversations and
the anonymous referee for suggesting model S1.  This research was
supported by a scholarship award from the Achievement Rewards for
College Scientists Foundation to S.D.R.; by NSF Career Grant
AST-0449986 and NASA Planetary Geology and Geophysics Program Grant
NNG04GK19G to G.L.; and by NSF Grant AST-0507424 and NASA Origins Grant
NNX08AH82G to P.B.

\begin{figure}
\plottwo{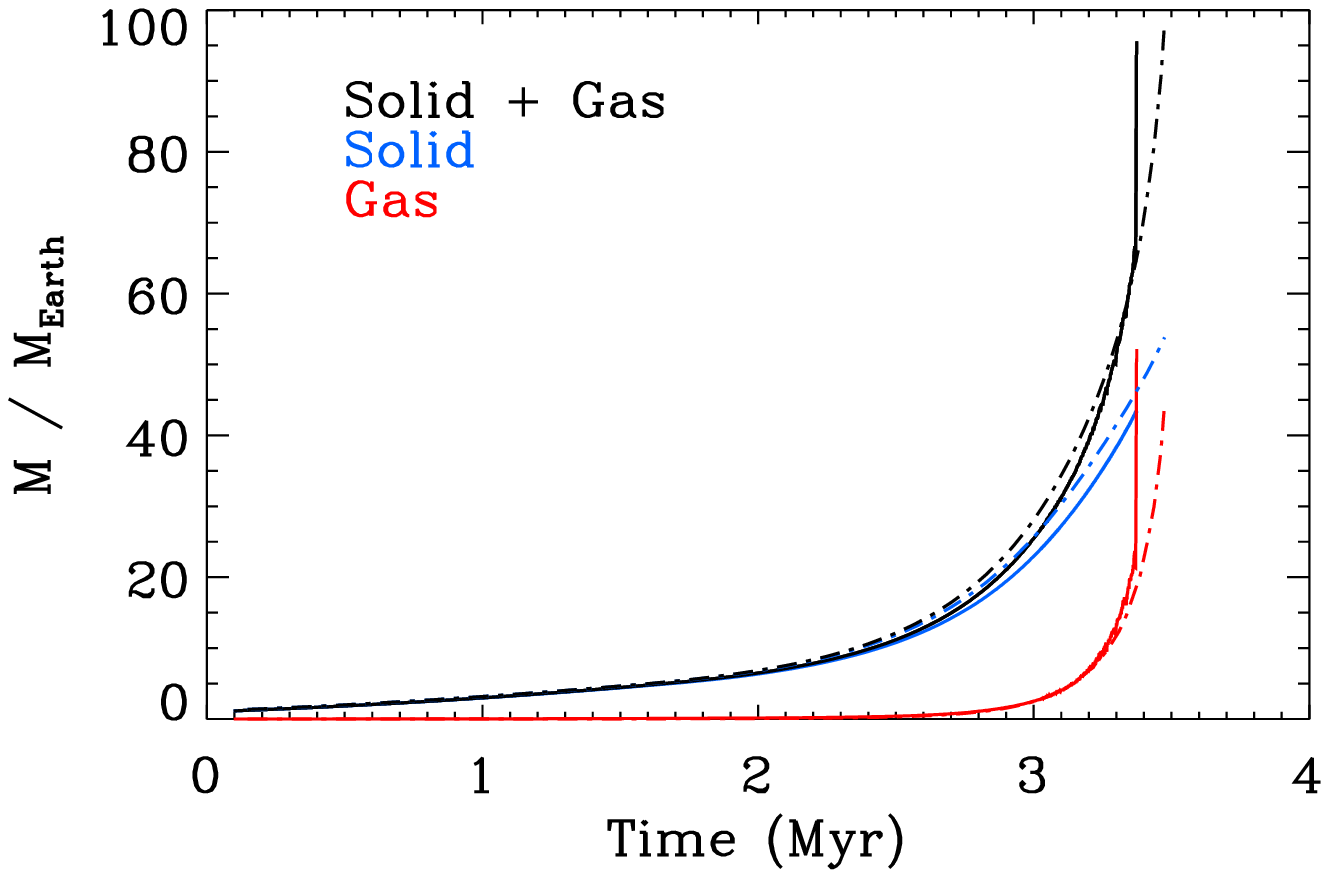}{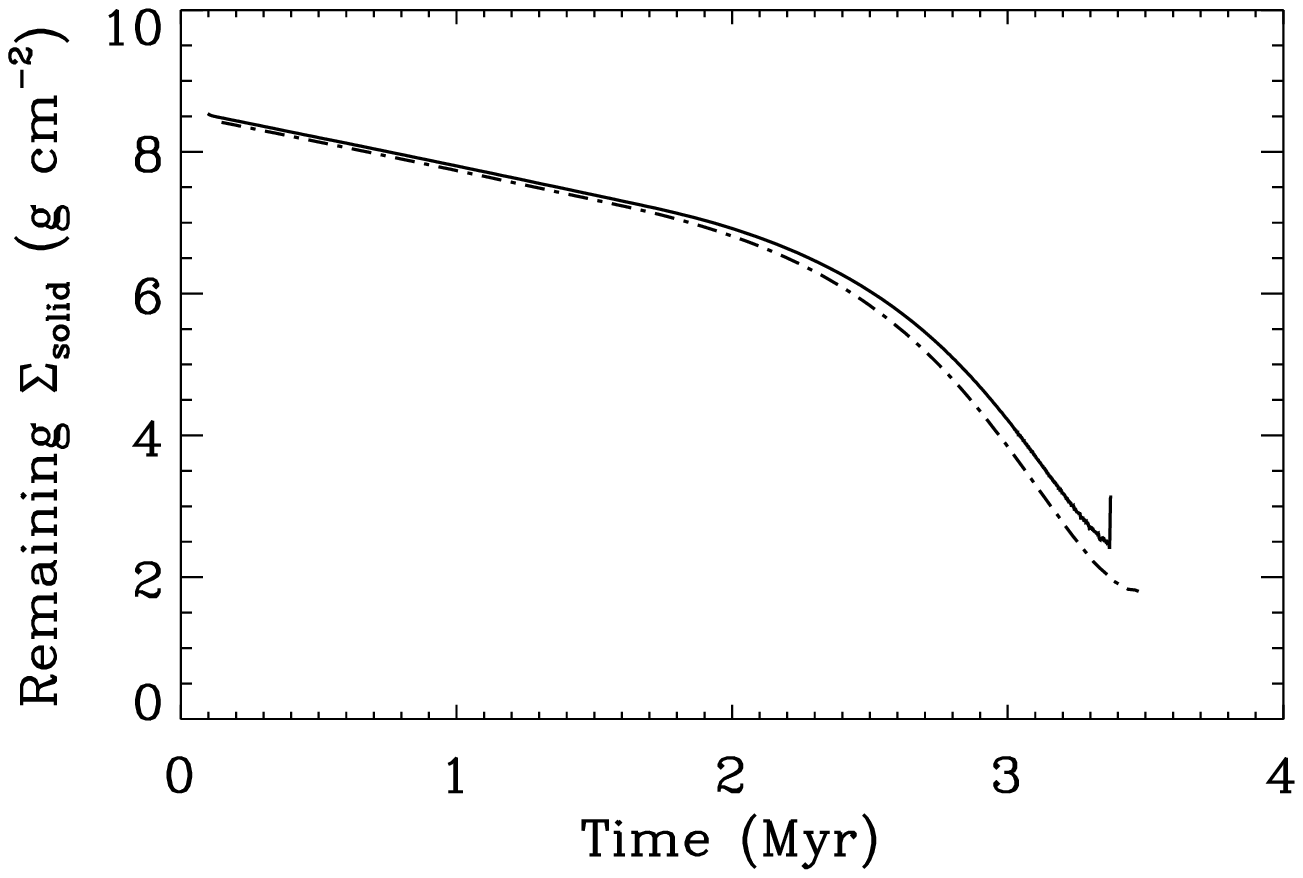}
\caption{{\bf a (Left):} Mass of Saturn as a function of time.  Solid
lines represent model S0 (reduced grain opacity) and dashed lines
represent model S1 (full grain opacity).  The black curves show the
total planet mass, the blue curves show the solid mass only (presumed to
be concentrated in the core), and the red curves show the gas mass.
Saturn reaches its current mass, $95 M_{\oplus}$, in 3.4~Myr.
{\bf b (Right):} Remaining solid surface density, not yet accreted
by Saturn, as a function of time.  Solid line shows model S0 and dashed
line shows model S1.  The sharp upturn at 3.3~Myr in S0 is due to
the rapid expansion of Saturn's feeding zone when hydrodynamic gas
accretion begins.}
\label{planetforms}
\end{figure}



\end{document}